\documentclass[twocolumn,showpacs,preprintnumbers,amsmath,amssymb]{revtex4}
\usepackage{epsfig}
\usepackage{dcolumn}
\usepackage{bm}
\usepackage{graphicx}

\begin{document}

\bibliographystyle{unsrt}

\title{Forcing function control of Faraday wave instabilities in viscous
shallow fluids}
\author{ Cristi\'an Huepe (1), Yu Ding (2), Paul Umbanhowar (2) and
 Mary Silber (1)}
\affiliation{(1) Department of Engineering Sciences and Applied
 Mathematics. \\ (2) Department of Physics and Astronomy. Northwestern
 University, 2145 Sheridan Road, Evanston, IL 60208-3112, USA}
\date{\today}


\begin{abstract}
We investigate the relationship between the linear surface wave
instabilities of a shallow viscous fluid layer and the shape of the
periodic, parametric-forcing function (describing the vertical
acceleration of the fluid container) that excites them. 
We find numerically that the envelope of the resonance tongues can
only develop multiple minima when the forcing function has more than
two local extrema per cycle.
With this insight, we construct a multi-frequency forcing function
that generates at onset a non-trivial harmonic instability which is
distinct from a subharmonic response to any of its frequency
components.
We measure the corresponding surface patterns experimentally 
and verify that small changes in the forcing waveform cause a 
transition, through a bicritical point, from the predicted 
harmonic short-wavelength pattern to a much larger standard 
subharmonic pattern.
Using a formulation valid in the lubrication regime (thin viscous
fluid layer) and a WKB method to find its analytic solutions,
we explore the origin of the observed relation between the forcing 
function shape and the resonance tongue structure.
In particular, we show that for square and triangular forcing
functions the envelope of these tongues has only one minimum, 
as in the usual sinusoidal case.
\end{abstract}

\pacs{47.35.+i,47.20.-k,47.54.+r}

\maketitle


\section{Introduction}

In the Faraday system, an incompressible fluid is oscillated
vertically in a container with a free upper surface, generating
standing surface waves which provide an excellent system for the study
of pattern formation \cite{Miles,Cross}.
Through an appropriate choice of experimental parameters, many of the
regular patterns that are possible in two dimensions, such as stripes,
squares and hexagons, can be obtained.
In addition, targets, spirals, superlattices and quasipatterns lacking
strict translational periodicity have also been observed
\cite{Christiansen1,Christiansen2,Kudrolli1,Wagner,Binks}.

One of the advantages of the Faraday experiment, when compared to
other pattern-forming systems such as convection or chemical
reactions, is the great amount of control over the energy feeding
mechanism that can be achieved by changing the periodic vertical
acceleration of the fluid container.
Even by forcing the system with different combinations of only two
frequencies, several distinct patterns can be achieved.
Hexagonal and rhomboid patterns, together with various quasipatterns
have been obtained experimentally in \cite{Fauve1,Fauve2,Arbell1} by
varying the amplitudes and the phase difference between both
components.
Superlattice patterns \cite{Kudrolli2,Arbell2}, triangular patterns
\cite{Muller1} and localized structures \cite{Arbell3} have also been
observed using two-frequency forcings \cite{Arbell4}.

From a theoretical perspective, a combination of tools must be used to
understand and predict the pattern selection.
While its characteristic wavelength can be obtained through a linear
instability calculation, the two-dimensional structure is determined
by the nonlinear interaction between modes
\cite{Zhang,Chen1,Silber0,Chen2,Silber1,Silber2,Silber3,Silber4,Silber5}.
At the linear level, the simplest cases occur when a deep fluid layer
of low viscosity is oscillated with a sinusoidal forcing, i.e.
proportional to $\sin(\omega t)$.
In these situations, the frequency of the main (largest in amplitude)
component of the resulting surface wave oscillations will be
$\omega/2$ (referred to hereafter as the first -or fundamental-
subharmonic response).
In other cases, two mechanisms for selecting the main frequency 
responses that are different from the first subharmonic one have been
identified.

The first mechanism occurs when two or more frequency components are
introduced in the forcing.
In these cases, each component will tend to excite its own
corresponding first subharmonic mode.
Their relative amplitudes will determine which of these responses has
the lowest global forcing strength threshold, thus becoming the
instability that is observed at onset.
%
%
The second mechanism can only arise in the high viscosity regime.
If the fluid layer is shallow enough, even a single component forcing
with low enough frequency can excite an instability different from the
first subharmonic one.
As the viscous boundary layer reaches the bottom of the fluid
container, the threshold of the lowest unstable modes rises, allowing
others with higher main frequency components (and, therefore, shorter
surface wavelengths) to become unstable at onset \cite{Kumar,Muller2}.
%

%
In a numerical and experimental study, it was shown in
\cite{Tuckerman2} that a transition between two patterns with
different linearly unstable wavelengths can be obtained in various
fluid regimes by changing the relative amplitudes of a two-frequency
forcing function.
This transition occurs through a bicritical point, where both modes
are simultaneously neutrally stable.
In spite of these results, only a limited understanding of the effects
of both a multi-frequency forcing and a high viscosity regime has been
achieved.
Furthermore, little is known about the patterns expected for more
complicated forcing functions not described by a few frequency
components.
This can be attributed to the essentially infinite number of degrees
of freedom that are needed to parametrize an arbitrary forcing
function, which renders a systematic exploration of the parameter
space impossible.

In this paper, we consider a different and novel approach. Instead of
exploring a large parameter space with various forcing frequency
components, we seek to identify which characteristics of the periodic
forcing function affect the surface patterns and how.
By performing a numerical linear stability calculation in various test
systems of shallow viscous fluid layers, we will first identify a
simple qualitative relation between the shape of the forcing function
and the resonance tongue structure (that describes the stability
thresholds).
Using this relation, we will construct a forcing function with a
non-trivial critical instability at onset, having a main frequency
component which does not correspond to the fundamental subharmonic 
(or even the fundamental harmonic) response to any of its forcing 
frequencies.
We will then present experimental results showing the surface pattern
generated by this instability.
%
%
Finally, in the lubrication limit of a thin viscous fluid layer, we
will illustrate analytically the origin of the observed relation
between the forcing function and the stability thresholds.
We will follow the method introduced by Cerda and Tirapegui
\cite{Cerda1,Cerda2} that derives a Mathieu equation to describe this
regime and uses a WKB approximation \cite{Messiah,Goldman} to solve it
for single frequency forcing.
By extending these calculations to arbitrary forcing functions we will
develop an intuitive understanding of the relation between the shape
of the forcing function and the structure of the resonance tongues. In
particular, we will show that only forcing functions with more than
two local extrema per cycle are expected to allow bicritical points
involving non-contiguous tongues.

The paper is organized as follows. In Section \ref{sec:Background} we
review the standard formulation of the Faraday wave linear stability
analysis.
We introduce in Section \ref{sec:Numerical} a one-parameter family of 
forcing functions to illustrate numerically the
relation between the shape of each member of the family and the
structure of its corresponding neutral stability diagram.
Section \ref{sec:Experimental} presents an experimental study that
uses these forcing functions, displaying a previously unobserved
transition between two surface patterns with very different
characteristic wavelengths.
In Section \ref{sec:Analytical} we show an approximate analytical
relation between the forcing and the instability response that
illuminates our approach.
Finally, Section \ref{sec:DiscAndConcl} briefly discusses our results
and presents our conclusions.
%


\section{Background}
\label{sec:Background}

We study the linear stability of the free surface of an incompressible
Newtonian fluid layer of depth $h$, density $\rho$, kinematic
viscosity $\nu$ and surface tension $\sigma$. The fluid is oscillated
vertically with acceleration $f(\omega t)$, where $\omega$ is the
fundamental frequency of oscillation and $t$ is the time.
We will summarize here the derivation of the equations describing this
system by following the presentation in \cite{Tuckerman1}.

Using the incompressibility condition to eliminate the pressure in the
linearized Navier-Stokes equation we obtain
\begin{equation}
\label{eq:linNS0}
(\partial_t - \nu \nabla^2) \nabla^2 u_z = 0,
\end{equation}
where $u_z(x,y,z,t)$ is the vertical component of the fluid velocity. 
In an idealized laterally infinite container, the horizontal
eigenfunctions are given by $e^{\pm i \vec{k} \cdot \vec{r}}$,
with $\vec{r} = (x,y)$ and $\vec{k} = (k_x,k_y)$.
For each surface wavenumber $k = |\vec{k}|$, equation (\ref{eq:linNS0})
thus becomes
\begin{equation}
\label{eq:linNS}
\left[\partial_t - \nu (\partial_{zz} - k^2) \right] 
(\partial_{zz} - k^2) v_k = 0,
\end{equation}
where $v_k(z,t)$ describes the $z$-dependence of $u_z$ associated 
with the mode $k$.
In the oscillating reference frame with $z=0$ at the flat fluid
surface, the boundary conditions on the bottom of the container are
given by
\begin{equation}
\label{eq:bc1y2}
v_k = 0 \ \,  \mbox{  and   } \,
\partial_z v_k = 0,  \, \, \mathrm{at} \, \, z = -h.
\end{equation}
At the fluid surface, the vertical position of the free boundary $z =
\xi_k(t) e^{i \vec{k} \cdot \vec{r}}$ associated to every mode $k$ is
advected by the fluid motion. To linear order in the surface
deformation this kinematic boundary condition is
\begin{equation}
\label{eq:bc3}
\frac{d \xi_k}{d t} = v_k \, \, \, \mathrm{at} \, \, z = 0.
\end{equation}
Additional boundary conditions are imposed at the surface by finding
the total balance of forces tangential and normal to the interface.
From this we obtain
\begin{eqnarray}
\label{eq:bc4}
\left( \partial_{zz} + k^2 \right) v_k = 0,&&  \\
\label{eq:bc5}
\left[ \partial_t - \nu ( \partial_{zz} - k^2 ) + 2 \nu k^2 \right]
\partial_z v_k =&&  \nonumber \\
\big[ g \left( 1 + \Gamma f(\omega t) \right) 
&+& \frac{\rho}{\sigma} k^2 \big] k^2 \xi_k, \\
&& \hspace{0.5cm} \mathrm{at} \ \ z = 0, \nonumber
\end{eqnarray}
where $g$ is the gravitational acceleration and $f$ is a
non-dimensional function defined to have $\max(|f(\omega t)|) =
1$. Therefore, $\Gamma$ corresponds to the maximum acceleration of the
forcing function, expressed in units of $g$.

Equation (\ref{eq:linNS}) and boundary conditions (\ref{eq:bc1y2})
through (\ref{eq:bc5}) fully describe the dynamics of the system.
Instead of integrating them directly, our numerical analysis will
focus on finding the stability threshold $\Gamma_c(k)$ given by the
critical value of $\Gamma$ at which the wavenumber $k$ becomes
unstable.


\section{Numerical study}
\label{sec:Numerical}


\subsection{Method}

We are interested in finding numerically the neutral stability curve
$\Gamma_c(k)$ for various forcing functions.
With this objective, we have extended the stability analysis method of
Kumar and Tuckerman \cite{Tuckerman1,Tuckerman2} to forcing functions
with an arbitrary number of frequency components.
In broad terms, this method consists first in expanding $v_k$ and
$\xi_k$ in a Floquet form
\begin{eqnarray}
v_k = e^{(\mu + i \phi)t} 
      \sum_j w_j(z) e^{i j \omega t} + \mathrm{c.c.}\\
\xi_k = e^{(\mu + i \phi)t} 
      \sum_j \zeta_j e^{i j \omega t} + \mathrm{c.c.}.
\end{eqnarray}
Here, $\mu + i \phi$ is the Floquet exponent, where we can set the
growth rate $\mu$ to $0$ to obtain marginal stability curves with
harmonic ($\phi = 0$) and subharmonic ($\phi = \omega/2$) temporal
responses.
Equations (\ref{eq:linNS})-(\ref{eq:bc4}) are then used to rewrite
(\ref{eq:bc5}) in the form
\begin{equation}
\label{Eq:Tuckerman1}
A_n \zeta_n = \Gamma \left[ f \zeta \right]_n,
\end{equation}
where $A_n$ is an algebraic function of the system parameters, which
does not depend on $f(\omega t)$, and $\left[ f \zeta \right]_n$ is the
$n$-th Fourier component of
\begin{equation}
f(\omega t) \, \sum_j \zeta_j e^{i j \omega t}.
\end{equation}
By introducing the explicit form of $f(\omega t)$, equation
(\ref{Eq:Tuckerman1}) can be expressed as an eigenvalue problem for
the forcing amplitude $\Gamma$ which can then be solved through
standard numerical techniques.
In order to extend the method to cases beyond the two-frequency
forcing computed in \cite{Tuckerman1,Tuckerman2}, we implemented this
algorithm in {\it Mathematica} \cite{Mathematica} and used the
program's symbolic algebra capabilities to automatically compute
$\left[ f \zeta \right]_n$ for any given $f(\omega t)$.
With this implementation, which is analogous to that presented in
\cite{Weizhong}, we are able to obtain efficiently the neutral
stability curves for any desired forcing function, regardless of its
frequency content.
%


\subsection{Results }

We restrict our study to shallow viscous fluid layers.
Since the specific value of the fluid constants within this regime
does not change our qualitative results or analysis, we will further
reduce the size of the parameter space by considering throughout the
paper only one set of fluid constants.
These are given by
a density $\rho = 0.95 \, \mathrm{g}/\mathrm{cm}^3$, 
a surface tension $\sigma = 20 \, \mathrm{dyn}/\mathrm{cm}$ 
and a viscosity $\nu = 46 \, \mathrm{cS}$.
Additionally, we will use in this section and in Section
\ref{sec:Experimental} a fluid depth $h=0.3 \, \mathrm{cm}$ and an
oscillation frequency $\omega = 2 \pi \, (10 \, \mathrm{Hz})$.

By using the numerical techniques described above, we explored the
structure of the marginal stability curves $\Gamma_c(k)$ for many
different $f(\omega t)$ including various piecewise constant,
piecewise linear, delta-like and multi-frequency functions.
While a precise characterization of how the features of $f(\omega t)$
correlate to those of $\Gamma_c(k)$ remains to be achieved, one of the
salient qualitative relations that we observed for all tested
functions is a connection between the extrema of $f(\omega t)$ and
the envelope of $\Gamma_c(k)$ that will be described below.
We will illustrate it here for a specific family of forcing functions,
which is the same as that in the experiments of Section
\ref{sec:Experimental}.

Consider the following set of forcing functions parametrized by $p$
\begin{equation}
\label{eq:forcingfunction}
f_p(\omega t) = \mathcal{N} \left[ 2.5 \cos(\omega t) + 3^p \cos(3
\omega t) - 5^p \cos(5 \omega t) \right],
\end{equation}
where $\omega$ is the fundamental frequency of oscillation and
$\mathcal{N}$ is a normalization constant which is defined so that
$\max(|f_p(\omega t)|) = 1$.
The specific form of (\ref{eq:forcingfunction}) is an arbitrary choice
which is not important for the qualitative behavior that we will focus
on here.
It was obtained by searching for a one-parameter family of forcing
functions that simultaneously includes members with a simple
triangular-like form ($p \approx -2$) and others that can produce
non-trivial surface-wave instabilities in an experimentally accessible
regime ($p \approx 1$).
%

\begin{figure}[t]
\centerline{\epsfxsize=8.5cm{\epsfbox{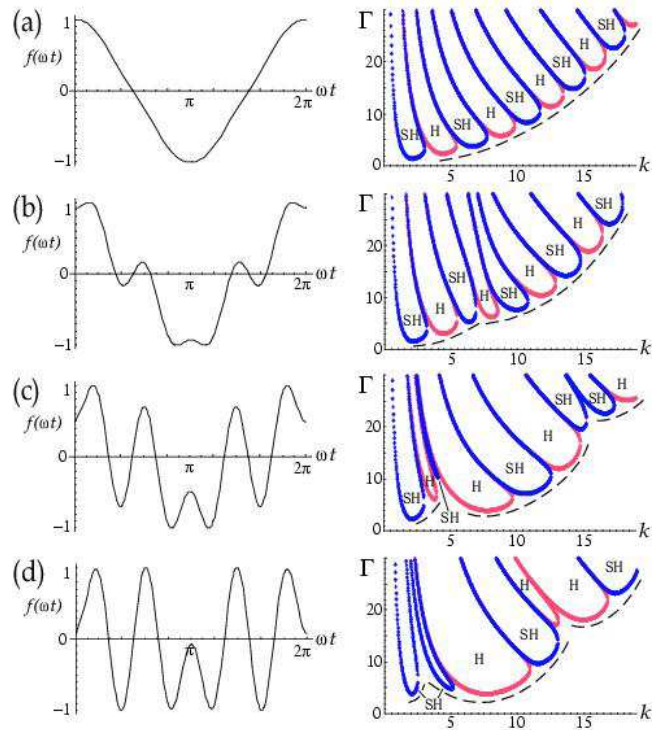}}} 
\caption{\label{fig:1} 
(Color online) Shape of the forcing functions (left) defined in
(\ref{eq:forcingfunction}) and their corresponding neutral stability
curves (right) for (a) $p=-2$, (b) $p=-0.3$, (c) $p=0.5$, (d) $p=1$,
and parameters $\rho = 0.95 \, \mathrm{g}/\mathrm{cm}^3$, $\sigma = 20
\, \mathrm{dyn}/\mathrm{cm}$, $\nu = 46 \, \mathrm{cS}$, $\omega = 2
\pi \, (10 \, \mathrm{Hz})$ and $h=0.3 \, \mathrm{cm}$.  $\Gamma$ is
in units of $g$ and $k$ in $\mathrm{cm}^{-1}$.  The resonance tongues
labeled H and SH show regions with harmonic or subharmonic linear
instabilities, respectively. Note how their envelopes (dashed lines)
change with $p$.}
\end{figure}

Figure \ref{fig:1} displays in the left column $f_p(\omega t)$ for $p
= -2$, $p = -0.3$, $p = 0.5$ and $p = 1$.
The right column shows the corresponding neutral stability curves
$\Gamma_c(k)$ which present the usual resonance tongue structure.  The
harmonic and subharmonic tongues indicate regions where surface waves
become unstable, oscillating with a main frequency component that is
an integral multiple ($ \omega, 2 \omega, 3 \omega, \ldots $) or an
odd half-multiple ($\omega/2, 3 \omega/2, 5 \omega/2, \ldots$) of the
fundamental forcing frequency, respectively.
The tongues at higher $k$-values correspond to instabilities with
shorter surface wavelengths and higher oscillation frequencies.
As $p$ is increased, the forcing function changes from a simple
rounded triangular shape with only two extrema per cycle to shapes
with richer structure.
Correspondingly, the envelope defined by the tongue minima (sketched
as a dashed line on the figure) changes from a simple convex function
with a single minimum to a set of convex segments, each with its own
minimum.

We have observed a similar relation between the structure of the
extrema of $f(\omega t)$ and the concavity of the resonance tongue's
envelope for all forcing functions tested (triangular, square,
multi-frequency, etc.)
In particular, every $f(\omega t)$ with only two extrema per cycle
resulted in an envelope with positive concavity for all $k$.
This relation will be one of our main focuses in the remainder of this
paper.
%

It is important to point out that the changes in the critical
instabilities illustrated in Fig.~\ref{fig:1} cannot be explained by a
simple switch to a different dominant forcing frequency in $f_p(\omega
t)$ combined with the first mechanism described in the Introduction.
Indeed, as $p$ is increased to $1$ the lowest unstable region becomes
the second harmonic tongue (with main frequency component equal to $2
\omega$) which does not correspond to the fundamental harmonic or
subharmonic responses (with equal or half the frequency, respectively)
to any of the three frequency components of $f_p(\omega t)$: $\omega$,
$3 \omega$ and $5 \omega$.
Furthermore, it is apparent that the change in $p$ cannot be
characterized as mainly reducing the stability threshold of a specific
tongue, but that it rather affects the aforementioned envelope over the
entire range of $k$ studied.


\section{Experimental Results}
\label{sec:Experimental}

In this section, we present experimental results showing that the
appearance of multiple minima in the envelope of the resonance tongues
can generate interesting measurable effects.
By carefully choosing the form of the forcing function, we find a
previously unobserved bicritical point between two surface patterns
with very different characteristic wavelengths.

In our experiments, we use silicone oil with 
$\rho = 0.95 \, \mathrm{g}/\mathrm{cm}^3$,
$\sigma = 20 \, \mathrm{dyn}/\mathrm{cm}$ and 
$\nu = 46 \, \mathrm{cS}$ (Fluka Silicone Oil AR 20), which are
the same fluid parameters as in Section \ref{sec:Numerical}.
A $0.3 \, \mathrm{cm}$ deep layer of this silicone oil is contained in
a cylindrical cell with a radius of $7.0 \, \mathrm{cm}$ and height of
$4.0 \, \mathrm{cm}$.
The cell has a PVC sidewall, a $0.8 \, \mathrm{cm}$ thick glass
bottom, and a $0.8 \, \mathrm{cm}$ thick plexiglass top covered with a
light diffuser.
It is mounted on the ram of a $10 \times 10 \, \mathrm{cm}^2$ linear
air bearing, which is attached to a 180 kg triangular granite slab
that floats on an air table to minimize horizontal oscillations.
A shaker (VTS VG100) is suspended by springs from the air table
supports.
Two $50 \, \mathrm{cm}$ long cylindrical aluminum tubes, each with an
inner and outer diameter of $0.48 \, \mathrm{cm}$ and $0.95 \,
\mathrm{cm}$, respectively, connect the shaker to the ram.
An amplifier (Crown CE2000) drives the shaker with a
computer-generated forcing function.
The amplitudes and phases of the desired Fourier components of the
acceleration signal are measured by an accelerometer (PCB Model
353B68) and used as feedback to control the driving.
The root-mean-square difference between the measured and target
forcing functions is less than 1\% while the variation in the
amplitudes of the driven components is less than 0.01\%.
Since viscosity and surface tension are both sensitive to temperature
changes, the experiments are conducted in a closed transparent box
maintained at a constant temperature ($\pm0.005^{\circ}$C).
To visualize the waves, parallel light is projected through the cell
bottom. The curved fluid surface refracts the light, which then falls
on the diffuser producing a representation of the pattern. 
A CCD camera synchronized with the forcing function acquires the
images.

Our specific choice of forcing function was determined by searching
for an experimentally achievable set of parameters having a linear
instability at onset with a response far from the usual subharmonic
one.
This objective is not easily achieved despite the fact that our
numerical exploration established that many forcing functions generate
resonance tongues with a multiple minima envelope.
Indeed, for the fluid parameters used in our experiments, we found
numerically that a tongue belonging to the second or higher (in order
of increasing $k$) envelope minimum can be excited at onset only for
very low values of $h$ or $\omega$.
However, the range of these two quantities is limited by our
experimental apparatus.
For very shallow fluid layers ($h < 0.1 \, \mathrm{cm}$), spurious
effects can affect the patterns: surface waves may contact the bottom
of the container and a small tilt, variation in the bottom profile, or
wetting at the wall can lead to large changes in the relative fluid
depth $\Delta h / h$.
Additionally, as $h$ and $\omega$ are reduced, the critical
acceleration $\Gamma_c$ increases.
Because the maximum acceleration and amplitude ($\propto
\omega^{-2}$) of the apparatus are limited, much of this low
$\omega$/large $\Gamma$ regime is inaccessible.
By testing numerically various forcing functions, we were able to
construct $f_p(\omega t)$ with $p \approx 1$, as defined in
(\ref{eq:forcingfunction}), which has a global minimum in the part of
the envelope that does not contain the first subharmonic tongue (see
Fig.~\ref{fig:1}d), and which is experimentally accessible.

\begin{figure}[t]
\centerline{\epsfxsize=8.5cm{\epsfbox{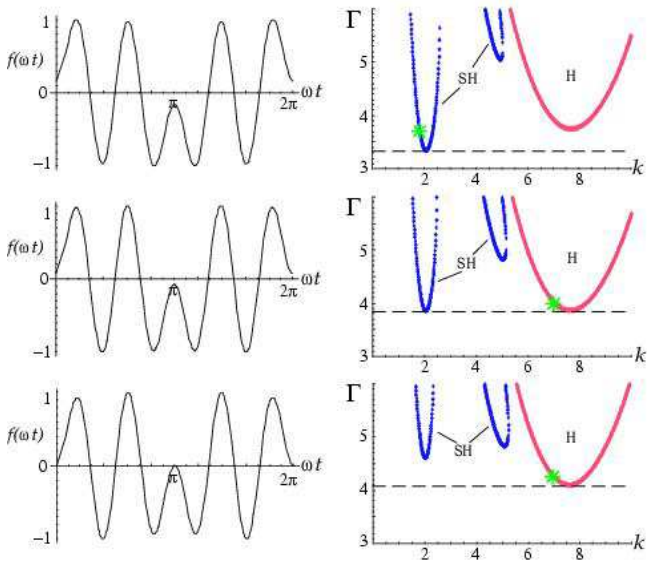}}} 
\caption{\label{fig:2}
(Color online) Shape of the forcing functions (left) and their 
corresponding neutral
stability curves (right) for $p=0.9$ (top), $p=1.0$ (center), $p=1.1$
(bottom), and the same parameters and units as in
Fig.~\ref{fig:1}. For small changes in $f(\omega t)$, the instability
at onset (occurring at critical forcings indicated by the dashed
lines) switches from the first subharmonic (SH) resonance tongue to
the second harmonic (H) one.
Each asterisk indicates the critical forcing $\Gamma_c$ and wavenumber
$k_c$ measured experimentally.
}
\end{figure}

Figure \ref{fig:2} displays the neutral stability curves computed
numerically for the experimental parameters specified above, using
$\omega = 2 \pi \, (10 \, \mathrm{Hz})$ and a forcing $f_p(\omega t)$
with $p=0.9$, $p=1.0$ and $p=1.1$.
%
%
The figure shows a very small change in the forcing function (see left
panels) producing a large jump in the critical wavenumber.
For $p = 0.9$ (top), the first subharmonic tongue (with main frequency
component at $\omega / 2$) will be excited at onset.
Numerically, we compute a critical forcing $\Gamma^{\mathrm{SH}}_c =
3.35$ and a critical wavenumber $k^{\mathrm{SH}}_c = 2.07$.
At $p = 1.0$ (center), the system is close to a bicritical point,
where the first subharmonic and second harmonic tongues become
simultaneously unstable at onset.
The corresponding critical values are
$\Gamma^{\mathrm{SH}}_c = 3.86$, $k^{\mathrm{SH}}_c = 2.05$ and
$\Gamma^{\mathrm{H}}_c = 3.87$, $k^{\mathrm{H}}_c = 7.64$,
respectively.
Finally, for $p = 1.1$ (bottom) the second harmonic tongue becomes the
instability at onset, with $\Gamma^{\mathrm{H}}_c = 4.10$ and
$k^{\mathrm{H}}_c = 7.59$.
We refer to it as the second harmonic one since it oscillates with a
main frequency component at $2 \omega$, and is therefore the second
harmonic tongue in order of growing $k$ (the first being above the
plotted $\Gamma$-range, between the two subharmonic tongues displayed).
It is a non-trivial critical instability, which cannot be easily
explained by the mechanisms described in the Introduction, because it
does not correspond to the first harmonic or subharmonic responses to
any of the forcing frequency components ($\omega$, $3 \omega$ and $5
\omega$). Instead, it is related to the second local minimum of the
envelope of the resonance tongues (see Fig.~\ref{fig:1}d).
%

\begin{figure}[t]
\centerline{\epsfxsize=8.5cm{\epsfbox{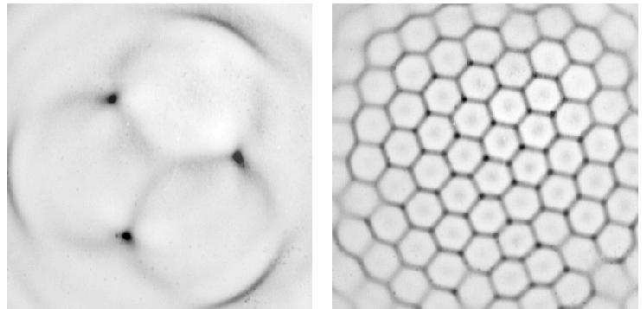}}} 
\caption{\label{fig:3} 
Experimental pictures (negative-images) of the surface patterns
appearing at onset for $p=0.9$ (left) and $p=1.1$ (right),
corresponding to the top and bottom forcing functions in
Fig.~\ref{fig:2}.
Note that, for this small variation in the forcing, a dramatic change
in the pattern is observed.
The size of each image is 
$8.22 \, \mathrm{cm} \times 8.22 \, \mathrm{cm}$,
which captures the central region of the container. 
}
\end{figure}

Experimentally, we observe the transition between these two linearly
unstable regimes by using the same $f_p(\omega t)$ forcing function.
Figure \ref{fig:3} shows images of the surface patterns for $p=0.9$
(left) and $p=1.1$ (right).
As predicted by our numerical calculations, their characteristic
length scale changes dramatically, in spite of the small variation in
$f_p(\omega t)$.
For $p=0.9$, we obtain a pattern of large hexagons at a critical
forcing $\Gamma_c = 3.72 g$, with a characteristic size of $3.5 \,
\mathrm{cm}$ which corresponds to the critical wavenumber $k_c = 1.77$.
%
%
When compared to the numerical results, $\Gamma_c$ is within $10 \%$
and $k_c$ within $ 15 \%$ of the predicted values.
Given the pattern deformation that is observed towards the image
borders due to the small aspect ratio (the size of the container is
only about twice the surface wavelength), these discrepancies are not
significant.
For $p=1.1$, a pattern of small hexagons appears at $\Gamma_c = 4.27$,
with a characteristic size of $0.9 \, \mathrm{cm}$, which implies $k_c
= 6.96$.
%
These measurements are within $4 \%$ (for $\Gamma_c$) and $9 \%$ (for
$k_c$) of the numerical predictions.
We have also verified in our experiments that, with respect to the
fundamental forcing frequency, the oscillations of the large pattern
are subharmonic and those of the small one are harmonic.
Finally, at $p=1.0$ (image not shown), we observe that the system
generates small hexagons which are practically indistinguishable from
those at $p=1.1$, with $\Gamma_c = 4.0$ and $k_c = 6.98$.

For $0.92 \leq p \leq 0.95$, we find in our experiments a bicritical
region where a complicated mixed mode surface pattern appears.
These kind of patterns can arise from the nonlinear interactions of
two or more linear instabilities
\cite{Silber1,Silber2,Silber3,Silber4,Silber5}.
They are often obtained by introducing frequency components in the
forcing function with simple linear responses that interact in the
horizontal plane to produce new structures.
In contrast, in the current situation the changes in the tongue
envelope selects linear instabilities that are not directly connected
to the forcing components, and therefore the patterns generated
through this mechanism could potentially be different.
Unfortunately, in our current experiment the mixed surface patterns
include complicated interactions with the side walls due to the small
size of the container.
Their proper analysis will therefore require a much larger aspect
ratio and is left for future work.
%


\section{Analytical Calculations}
\label{sec:Analytical}


\subsection{The lubrication approximation}

\begin{figure}[t]
\centerline{\epsfxsize=8cm{\epsfbox{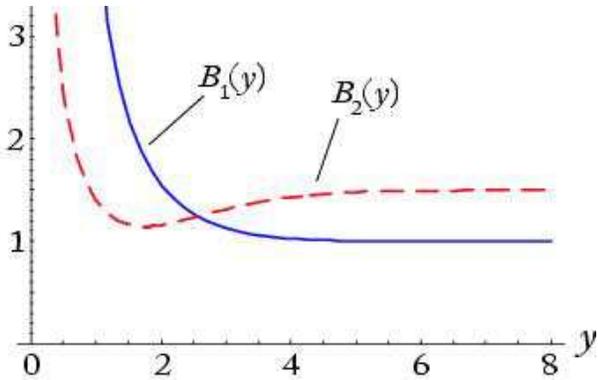}}} 
\caption{\label{fig:4} 
  (Color online) Plot of functions $B_1$ (solid line) and $B_2$
  (dashed line), as defined by equations (\ref{eq:B1}) and
  (\ref{eq:B2}). Both functions diverge for $y \rightarrow 0$ with
  $B_1(y) \propto y^{-3}$ and $B_2(y) \propto y^{-1}$. For $y
  \rightarrow \infty$ they approach their corresponding asymptotic
  limits $B_1(y) \rightarrow 1$ and $B_2(y) \rightarrow 3/2$.}
\end{figure}

We are interested in exploring analytically the origin of the relation
observed in Section \ref{sec:Numerical} between $f(\omega t)$ and the
envelope of $\Gamma_c(k)$.
To proceed, we will focus on systems in the {\em lubrication regime},
where the ratio between the $\partial_t \vec{v}$ and the $\nabla^2
\vec{v}$ term of the Navier-Stokes equation is small.
This ratio is of order $(l / \delta)^2$, where $l$ is the distance
that the fluid motion penetrates the surface and $\delta$ is the
characteristic size of the boundary layer \cite{Cerda1,Cerda2,Landau}.
Since $l$ can be estimated by either $1/k$ (if $k h \gg 1$) or $h$ (if
$k h \leq 1$), and $\delta$ is proportional to $\sqrt{\nu/\omega}$, 
it follows that a system is in the lubrication regime if it consists
of a shallow enough fluid layer with high enough viscosity and a low
enough oscillation frequency.

We use a simplified analytic description, introduced by Cerda and
Tirapegui in \cite{Cerda1,Cerda2} for fluids under the lubrication
approximation, in which a damped Mathieu equation involving only the
motion of the free fluid surface is obtained.
This equation is found by first deriving an exact non-local (in time)
relation for the linear evolution of the surface, which is a
formulation analogous to that developed in \cite{Beyer}.
By imposing a short-memory to the system due to its fast dissipation
rate, the non-local dependence is then neglected.
The resulting Mathieu equation reads
\begin{equation}
\label{eq:MathieuTC}
  \ddot{\xi}_k + 2 \bar{\gamma}_k \dot{\xi} + 
  \bar{\omega}^2_k \left[ 1 + \Gamma_k f(\omega t) \right] \xi_k = 0,
\end{equation}
where the dots represent derivatives with respect to time and
\begin{eqnarray}
\bar{\gamma}_k &=& \nu k^2 B_1(k h) B_2(k h) \\
\bar{\omega}^2_k &=& 
k \left[ g + \sigma k^2 / \rho \right] B_2(k h) \\
\label{def:Gammak}
\Gamma_k &=& \frac{\Gamma g}{ g + \sigma k^2 / \rho }.
\end{eqnarray}
Here, $B_1(k h)$ and $B_2(k h)$ are explicit non-dimensional functions
given by
\begin{eqnarray}
\label{eq:B1}
&B_1(y) = \frac{ \cosh(2 y) + 2 y^2 + 1}{ \sinh(2 y) - 2 y } &\\
\label{eq:B2}
&B_2(y) = \frac{ 3 \cosh^2(y) \left[ \sinh(2 y) - 2 y - 4 y^3 / 3 \right] + 
  y^2 \left[ \sinh(2 y) - 2 y \right] }{\left[ \sinh(2 y) - 2 y \right]^2}.&
\end{eqnarray}
Figure \ref{fig:4} shows that $B_1(y)$ and $B_2(y)$ have a simple
structure despite their complicated algebraic expressions. As $y$
approaches $0$, both functions diverge with $B_1(y) \propto y^{-3}$
and $B_2(y) \propto y^{-1}$. For large values of $y$, $B_1(y)$ and
$B_2(y)$ quickly converge to their asymptotic limits of $1$ and $3/2$,
respectively.

The critical forcing strength $\Gamma_c$ can be found for every $k$ by
considering solutions of (\ref{eq:MathieuTC}) that follow the Floquet
form
\begin{equation}
\label{eq:Floq1}
\xi_k(t + 2 \pi / \omega) = 
          e^{ \frac{2 \pi}{\omega} (\mu + i \phi) } \xi_k(t),
\end{equation}
and demanding that the growth rate after every period satisfies 
$\mu = 0$.


\subsection{The WKB approximation}

We will follow here the approach in \cite{Cerda1,Cerda2}, which uses
the well known (in the context of quantum mechanics)
Wentzel-Kramer-Brillouin (WKB) approximation \cite{Messiah,Goldman} to
solve the Mathieu equation.
We first cast (\ref{eq:MathieuTC}) into the form of a Schr\"odinger
equation by defining
\begin{eqnarray}
x &=& \omega t \\ 
\Psi(x) &=& \xi_k(x/\omega) e^{\bar{\gamma}_k x / \omega}
\end{eqnarray}
and
\begin{eqnarray}
E &=& \bar{\omega}^2_k - \bar{\gamma}^2_k \\
V(x) &=& -\Gamma_k \bar{\omega}^2_k f(x),
\end{eqnarray}
to obtain
\begin{equation}
\label{eq:Scrodinger}
\Psi''(x) + 
\frac{1}{\omega^2} \left[ E - V(x) \right] \Psi(x) = 0,
\end{equation}
where the double prime represents the second derivative with 
respect to $x$.
The problem of finding the solutions of (\ref{eq:MathieuTC}) that follow
the Floquet form (\ref{eq:Floq1}) then becomes equivalent to finding
the eigenfunctions of (\ref{eq:Scrodinger}) that satisfy
\begin{equation}
\label{eq:Floq2}
\Psi(x + 2 \pi) = e^{\frac{2 \pi}{\omega} 
         \left( \mu + i \phi + \bar{\gamma}_k \right) } \Psi(x),
\end{equation}
where the neutral stability curves are obtained for $\mu = 0.$
%

\begin{figure}[t]
\centerline{\epsfxsize=8.5cm{\epsfbox{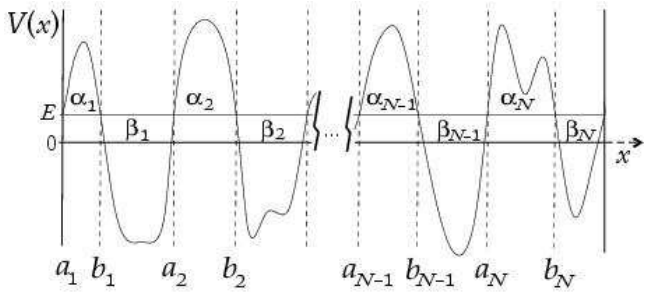}}} 
\caption{\label{fig:5} 
Illustration of the intervals of $V(x)$ in which the integrals
$\alpha_j$ and $\beta_j$ are computed using expressions
(\ref{eq:alphabeta1}) and (\ref{eq:alphabeta2}).
In the WKB approach described (see text), a matrix $M_j$ is defined
through expression (\ref{eq:Mn}) for each
$(\alpha_j,\alpha_{j+1})$-interval.  }
\end{figure}

In regions where $\omega^2/|E - V(x)| \ll 1$, the WKB approximation
provides explicit solutions for (\ref{eq:Scrodinger}) which are
divided into two different families.
For $E < V(x)$ (as in the $(a_j,b_j)$-intervals of Fig.~\ref{fig:5})
they are given in their most general form by
\begin{eqnarray}
\label{eq:PsiExp}
\lefteqn{\Psi(x) = \frac{1}{\sqrt{\omega P(x)}} \times} \nonumber \\
&& \left\{
  A \exp \left[ \int_{x_0}^x P(\tilde{x}) d\tilde{x} \right] +
  B \exp \left[ -\int_{x_0}^x P(\tilde{x}) d\tilde{x} \right]
\right\} \, \, \, \, \, \, \, \,
\end{eqnarray}
and for $E > V(x)$ (intervals $(b_j,a_{j+1})$ in Fig.~\ref{fig:5}), by 
\begin{eqnarray}
\label{eq:PsiSin}
\lefteqn{\Psi(x) = \frac{1}{\sqrt{\omega P(x)}} \times}  \nonumber \\
&& \left\{ 
  C \sin \left[ \int_{x_0}^x P(\tilde{x}) d\tilde{x} \right] +
  D \cos \left[ \int_{x_0}^x P(\tilde{x}) d\tilde{x} \right]
\right\}. \, \, \,
\end{eqnarray}
Here, $P(x) = \sqrt{|E - V(x)|} / \omega$ and the complex constants
$A$, $B$, $C$ and $D$ are obtained by imposing the boundary
conditions in each segment.
The solution for a given $V(x)$ over the full $x$ domain is found by
matching adjacent segments of $\Psi(x)$ at the points $x_0$
where $V(x_0) = E$.
For $x \approx x_0$, however, expressions (\ref{eq:PsiExp}) and
(\ref{eq:PsiSin}) are not valid and, following the WKB method, one
must perform a matched asymptotic expansion around $x_0$ to find the
correct matching formulas \cite{Goldman}.
At the points $\{ b_j \}_{j=1 \ldots N}$ shown on Fig.~\ref{fig:5},
these are given by
\begin{eqnarray}
\label{eq:match1}
C = \frac{2 A - B}{\sqrt{2}} \ \, \, \mbox{   and   } \, \,
D = \frac{2 A + B}{\sqrt{2}},
\end{eqnarray}
and at the points $\{ a_j \}_{j=1 \ldots N}$, by
\begin{eqnarray}
\label{eq:match2}
A = \frac{C + D}{\sqrt{2}} \ \, \, \mbox{   and   } \, \,
B = \frac{D - C}{2 \sqrt{2}}.
\end{eqnarray}

We will now extend the neutral stability calculations carried out in
\cite{Cerda1,Cerda2} for $V(x) \propto \cos(x)$ to arbitrary forcing
functions.
Imagine a periodic function $V(x)$ with $2 N$ matching points per
period as in Fig.~\ref{fig:5}.
Using (\ref{eq:match1}) and (\ref{eq:match2}) we can relate the
coefficients $A_{j+1}$ and $B_{j+1}$ of solution (\ref{eq:PsiExp}) in
an interval $( a_{j+1},b_{j+1} )$ to the coefficients $A_{j}$ and
$B_{j}$ in the previous interval $( a_{j},b_{j} )$ (see
Fig.~\ref{fig:5}).
We find
\begin{equation}
\left(
\begin{array}{c}
A_{j+1} \\
B_{j+1}
\end{array}
\right)
= M_j
\left(
\begin{array}{c}
A_{j} \\
B_{j}
\end{array}
\right),
\end{equation}
where the matrix $M_j$ is defined by
\begin{equation}
\label{eq:Mn}
M_j = \left[
\begin{array}{ c c }
2 e^{\alpha_j} \cos(\beta_{j}) & 
                        -e^{-\alpha_j} \sin(\beta_{j}) \\
e^{\alpha_j} \sin(\beta_{j}) & 
                   \frac{1}{2} e^{-\alpha_j} \cos(\beta_{j})
\end{array}
\right],
\end{equation}
with
\begin{eqnarray}
\label{eq:alphabeta1}
\alpha_j &=& \int_{a_j}^{b_j} P(\tilde{x}) d\tilde{x}  \\
\label{eq:alphabeta2}
\beta_j  &=& \int_{b_j}^{a_{j+1}} P(\tilde{x}) d\tilde{x}.
\end{eqnarray}
The change in the amplitude of the wave function $\Psi(x)$ after a
full period is therefore given by the product $M = M_N M_{N-1} \ldots
M_1.$
Hence, for solutions with the Floquet form, equation (\ref{eq:Floq2})
implies the neutral stability condition
\begin{equation}
\label{eq:StabCrit1}
\max( | \lambda_+ | , | \lambda_- | ) =  
         e^{\frac{2 \pi}{\omega} \bar{\gamma}_k },
\end{equation}
where $\lambda_+$ and $\lambda_-$ are the two eigenvalues of $M$.
An equivalent condition can be found by using the fact that the trace
$\mathrm{Tr}(M)$ is real and that the determinant $\mathrm{Det}(M)$ is
equal to $1$, together with the standard relations $\mathrm{Tr}(M) =
\lambda_{+} + \lambda_{-}$ and $\mathrm{Det}(M) = \lambda_{+}
\lambda_{-}$. 
The resulting expression is 
\begin{equation}
\label{eq:StabCrit2}
\mathrm{Tr}(M) = 
     \pm 2 \cosh \left({\frac{2 \pi}{\omega} \bar{\gamma}_k } \right),
\end{equation}
where the plus or minus signs provide the neutral stability
boundaries for harmonic or subharmonic resonances, respectively.

Note that for some values of $k$ and $\Gamma$ it is also possible to
have $E > V(x)$ or $E < V(x)$ for all $x$, and therefore no
intersections between $V(x)$ and $E$.  In these situations the matrix
$M$ cannot be computed and our current implementation breaks
down. However, the WKB method is still valid and it has been shown in
\cite{Cerda1,Cerda2} that these cases never lead to instabilities.
In our computation of the neutral stability curves we can therefore
assume that there is at least one $\alpha$ and one $\beta$ region per
cycle.

\subsection{Validity of the approximation.}

\begin{figure}[t]
\centerline{\epsfxsize=8.8cm{\epsfbox{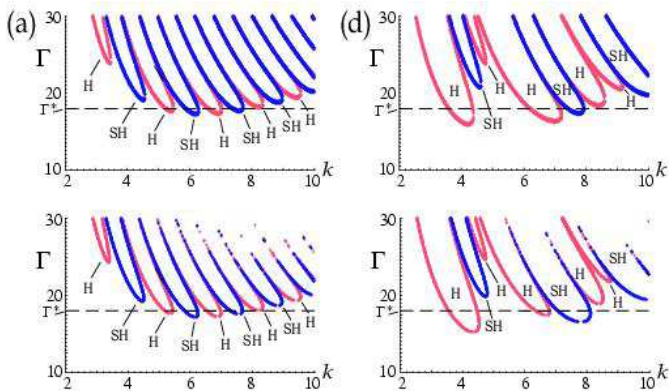}}} 
\caption{\label{fig:6} 
(Color online)
Neutral stability curves for a forcing function $f_p(\omega t)$ with
(a) $p=-2$, (d) $p=1$ (labeled as in Fig.~\ref{fig:1}) and parameters
$\rho = 0.95 \, \mathrm{g}/\mathrm{cm}^3$, $\sigma = 20 \,
\mathrm{dyn}/\mathrm{cm}$, $\nu = 46 \, \mathrm{cS}$, $\omega = 2 \pi
\, (3.5 \, \mathrm{Hz})$ and $h = 0.1 \, \mathrm{cm}$.
$\Gamma$ is in units of $g$ and $k$ in $\mathrm{cm}^{-1}$.
The exact numerical computations (top) are compared to the WKB
approximation (bottom).
The shape of the harmonic (H) and subharmonic (SH) resonance tongues
is essentially identical for $p=-2$ and has similar characteristics
for $p=1$.
In both cases, the tongues that would become unstable under a forcing
of $\Gamma^* = 18$ (dashed line) coincide.}
\end{figure}

We will investigate here the validity conditions for the approximation
described above.  The WKB method is based on an expansion in the small
quantity $\omega^2/|E - V(x)|$ which can be estimated by
\cite{Cerda1,Cerda2}
\begin{equation}
\label{eq:estimate}
\frac{\omega^2}{| E - V(x) |} \sim 
\frac{\omega^2}{\bar{\gamma}^2_k} \sim
\left( \frac{l}{\delta} \right)^4.
\end{equation}
This criterion implies that the approximation should be valid for
systems with $(l / \delta)^4 \ll 1$, which is a condition that must be
satisfied in the lubrication regime in which we are focusing. Indeed,
the lubrication regime requires $( l / \delta)^2 \ll 1$ and
therefore, given that $(l / \delta)^4$ will be even smaller, the WKB
approximation must also be valid in this regime.
Let us estimate $\delta$ and $l$ for the fluid parameters used in
Sections \ref{sec:Numerical} and \ref{sec:Experimental}.
For surface waves oscillating at a frequency $\Omega_k$, the
characteristic size $\delta$ of the viscous boundary layer is of order
$\sqrt{\nu / \Omega_k}$ \cite{Cerda1,Cerda2,Landau}.
Since the response frequency of the dominant surface waves is
typically of the same order as the forcing frequency, we have that
$\delta \sim \sqrt{ 0.46 / 10 } \approx 0.2 \, \mathrm{cm}$.
On the other hand, the distance $l$ that the motion of the surface
penetrates the fluid can be estimated by the smallest value between
$h=0.3 \, \mathrm{cm}$ and $1/k$.
In the region of $k$ considered (see Fig.~\ref{fig:6}), $l$ is
therefore larger than $\sim 0.1 \, \mathrm{cm}$.
Hence, for these parameters we have that $l / \delta$ is of order $1$,
which implies that the WKB method does not provide a good
approximation.

In order to be able to use a WKB analysis in our study, we will
consider in this section a shallower fluid layer with $h=0.1 \,
\mathrm{cm}$ and a lower oscillation frequency of $3.5 \,
\mathrm{Hz}$, while keeping all other parameters unchanged.
For this case, we have
$ \delta \sim \sqrt{ 0.46 / 3.5 } \approx 0.4,$
and $l \sim h = 0.1$. We thus obtain $(l / \delta)^4 < 10^{-2}$, which
should imply a good WKB approximation.
However, this criterion alone does not guarantee the accuracy of the
resulting neutral stability curves.
Indeed, for any forcing function there will be regions of $x$ where
$\omega^2 / | E - V(x) | \gg 1$, in which (\ref{eq:PsiExp}) and
(\ref{eq:PsiSin}) are not good approximations.
Unfortunately, the effect of these regions over the full periodic
$\Psi(x)$ solution cannot be easily estimated.
This problem becomes even harder if $V(x)$ has a complicated shape
because in such cases no simple approximation can even provide the
number or size of these regions, which depend on $k$ and $\Gamma$.
We will therefore validate our analysis by directly comparing the WKB
results to the numerical solutions of the full Navier-Stokes linear
stability problem.

Figure \ref{fig:6} shows the neutral stability curves obtained using
$\omega = 2 \pi \, (3.5 \, \mathrm{Hz})$, $h = 0.1 \, \mathrm{cm}$ and
the forcing function $f_p(\omega t)$ defined in expression
(\ref{eq:forcingfunction}) with $p = -2$ and $p = 1$ (labeled here `a'
and `d', as in Fig.~\ref{fig:1}).
The top panels show the exact numerical results computed using the
method described in Section \ref{sec:Numerical}, while the bottom ones
present the approximate WKB solutions.
The implementation of the WKB algorithm consists in finding the
values $\Gamma_c(k)$ for which the trace of $M$ satisfies
(\ref{eq:StabCrit2}), where $M$ is obtained by multiplying the
explicit expressions for $M_j$ given in (\ref{eq:Mn}).
By comparing the top and bottom panels, it is apparent that the WKB
curves are almost indistinguishable from the exact results in the 
$p = -2$ case. For $p = 1$, the WKB approximation and the exact
solution present a similar tongue structure but they do not coincide
in the exact predicted values for the critical stability threshold of
each tongue. However, the characteristics of their resonance tongue
envelopes are the same. This is the relevant feature here
since it is this envelope structure that we will study below
using the WKB method.


\subsection{Analysis of the envelopes}

\begin{figure}[t]
\centerline{\epsfxsize=7.5cm{\epsfbox{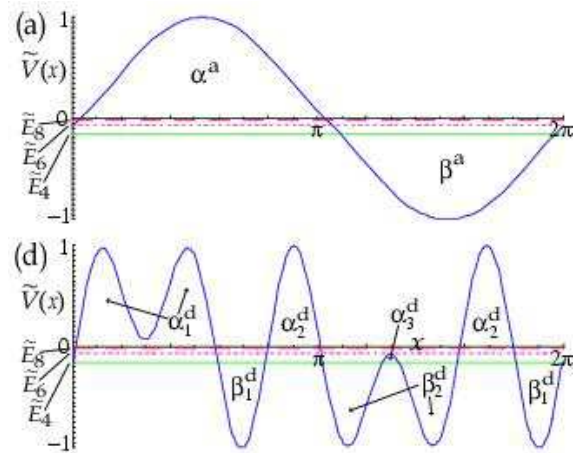}}} 
\caption{\label{fig:7} 
(Color online) Integration regions for the WKB calculations pertinent 
to Figs.~\ref{fig:8} and \ref{fig:9}.
The rescaled $\tilde{V}(x) = V(x)/(\Gamma_k \bar{\omega}^2_k)$ curves
correspond to a forcing $f_p(x)$ with (a) $p=-2$ and (d) $p=1$
(labeled as in Fig.~\ref{fig:1}).
The values of $\tilde{E}_k = E /(\Gamma_k \bar{\omega}^2_k)$ are
displayed for a forcing strength of $\Gamma^* = 18$ (see
Fig.~\ref{fig:6}) at $k=4$, $k=6$ and $k=8$.
Note that the integration zone $\alpha^d_3$ is not present for
$\tilde{E}_6$ and $\tilde{E}_8$ in (d).
}
\end{figure}

\begin{figure}[t]
\centerline{\epsfxsize=7.5cm{\epsfbox{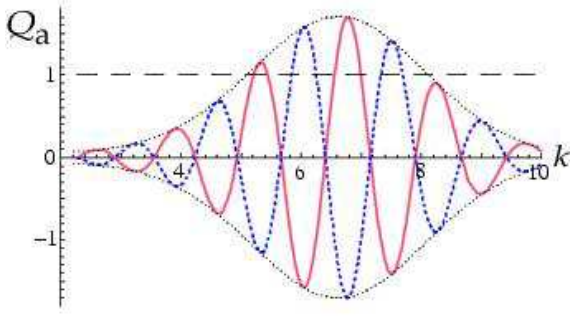}}} 
\caption{\label{fig:8} 
(Color online)
Plot of $Q_a(k,\Gamma^*)$ for a forcing function $f_p(\omega t)$ with
$p = -2$ and a forcing strength $\Gamma^*$ (see Fig.~\ref{fig:6}).
The regions with $Q_a > 1$ are unstable with harmonic (solid curve) or
subharmonic (dashed) responses.  The dotted envelope is computed by
discarding the $\cos(\beta^a)$ factor in equation (\ref{eq:QA}).
}
\end{figure}

Using the WKB approximation, we are now in a position to relate the
shape of the forcing function to the resonance tongue envelope.
For any $k$ and $\Gamma$, the stability criterion (\ref{eq:StabCrit2})
can be computed in terms of
\begin{equation}
\label{eq:Q}
Q(k,\Gamma)  \equiv  \pm \frac{\mathrm{Tr}(M)}
    {2 \cosh \left( 2 \pi \bar{\gamma}_k / \omega \right)},
\end{equation}
where $Q(k,\Gamma) > 1$ indicates an instability.
If the forcing function has only two extrema per cycle, there will
always be at most one $\alpha$ and one $\beta$ integration region,
as illustrated on Fig.~\ref{fig:7} (top) for $f_p(\omega t)$ with
$p=-2$ (labeled by an `a', as in Figs.~\ref{fig:1} and \ref{fig:6}).
In these cases we have $M = M_1$, and (\ref{eq:Q}) becomes
\begin{equation}
\label{eq:QA}
Q_a(k,\Gamma) = \pm 
\frac{\cosh( \alpha^a + \log 2 ) \cos(\beta^a)}{
  \cosh \left( 2 \pi \bar{\gamma}_k / \omega \right)}.
\end{equation}
If we consider the function $Q_a(k)$ at constant $\Gamma$, the
$\cos(\beta^a)$ factor will be responsible for oscillations that
generate an unstable tongue at every excursion that reaches $Q_a>1$.
Figure \ref{fig:8} plots $Q_a$ at a fixed forcing strength 
$\Gamma^*=18 \, g$,
indicated by the dashed horizontal line on Fig.~\ref{fig:6}.
The dotted lines trace the envelope of $Q_a$, which is readily
obtained by discarding the $\cos(\beta^a)$ factor from (\ref{eq:QA}).
It exhibits a single maximum on the figure and for all other values of
$\Gamma$ tested, implying that the envelope of the resonance tongues
must have a single minimum.
%

\begin{figure}[t]
\centerline{\epsfxsize=7.5cm{\epsfbox{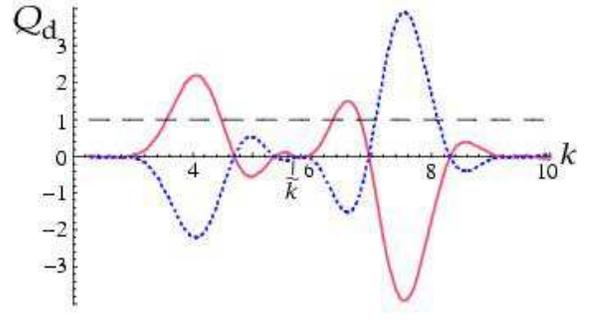}}} 
\caption{\label{fig:9}
(Color online)
Plot of $Q_d(k,\Gamma^*)$ for a forcing function $f_p(\omega t)$ with
$p = 1$ and a forcing strength $\Gamma^*$ (see Fig.~\ref{fig:6}).
The regions with $Q_d > 1$ are unstable with harmonic (solid curve) or
subharmonic (dashed) responses.
At $\tilde{k}$ the definition of $Q_d$ switches from $Q^{<}_d$ to
$Q^{>}_d$, given by Eqs. (\ref{eq:QD<}) and (\ref{eq:QD>}),
respectively, since the integration region $\alpha^d_3$ is not present
for $k > \tilde{k}$ (see Fig.~\ref{fig:7}d).
}
\end{figure}

In contrast, forcing functions with multiple extrema produce more
complicated envelope structures.
Figure \ref{fig:9} shows a plot of $Q(k,\Gamma^*)$ for $f_p(\omega t)$
with $p = 1$ (labeled here $Q_d$ since it corresponds to case `d'
in Figs.~\ref{fig:1}, \ref{fig:6} and ~\ref{fig:7}).
The oscillation amplitude presents two distinct zones of local maxima
at $k \sim 4$ and $k \sim 7$, which are responsible for the two minima
that the envelope of the resonance tongues displays in
Fig.~\ref{fig:6}.
In general, it is easy to see that any resonance tongue envelope with
multiple minima must be associated with $Q(k)$ functions (at fixed
$\Gamma$ values) which have amplitude envelopes with multiple maxima.
We will now study how these complicated amplitude envelopes arise by
examining in detail the analytical form of $Q_d$.
%

The bottom panel of Fig.~\ref{fig:7} shows the integration regions
for the $p = 1$ case.
Here, $M$ is given by the product of either three or four matrices,
depending on the $k$-interval considered, since the $\alpha_3$ region
is present for $k < \tilde{k} \approx 5.7$, but not for $k >
\tilde{k}$.
In the $k < \tilde{k}$ case it is straightforward to compute that
\begin{equation}
\label{eq:QD<}
Q_d^{<}(k,\Gamma^*) \approx
H_C^{<} \ C_1^{<} \ C_2^{<} + H_S^{<} \ S_1^{<} \ S_2^{<},
\end{equation}
with
\begin{eqnarray}
H_C^{<}(k) &=& \frac{\cosh(\alpha_1^d + 2 \alpha_2^d + \alpha_3^d + 
\log 16)}{\cosh ( 2 \pi \bar{\gamma}_k / \omega )}\\
H_S^{<}(k) &=& \frac{\cosh(\alpha_1^d + 2 \alpha_2^d - \alpha_3^d + 
\log 4)}{\cosh ( 2 \pi \bar{\gamma}_k / \omega )},
\end{eqnarray}
and
\begin{eqnarray}
C_1^{<}(k) &=& \cos^2 \left( \beta_1^d \right) \ \hspace{0.7cm} \ 
 C_2^{<}(k) = \cos^2 \left( \frac{\beta_2^d}{2} \right) \\
S_1^{<}(k) &=& -\sin^2 \left( \beta_1^d \right) \ \hspace{0.3cm} \ 
 S_2^{<}(k) = -\sin^2 \left( \frac{\beta_2^d}{2} \right).
\end{eqnarray}
In (\ref{eq:QD<}), we have neglected several additional terms of a
similar form, but where the argument of the hyperbolic cosine
contained $-\alpha_1^d$ or $-\alpha_2^d$ contributions.
These terms turn out to be negligible when compared to $H_C^{<}(k)$
and $H_S^{<}(k)$ since $\alpha_1$ and $\alpha_2$ are of the same
order, and are much larger than $\alpha_3$ (see Fig.~\ref{fig:7}d).

For $k >\tilde{k}$, $M$ is composed of the product of only three
matrices and the expressions become simpler. Using an equivalent
approximation we obtain
\begin{equation}
\label{eq:QD>}
Q_d^{>}(k,\Gamma^*) \approx H_C^{>} \ C_1^{>} \ C_2^{>},
\end{equation}
with
\begin{eqnarray}
H_C^{>}(k) &=& \frac{\cosh(\alpha_1^d + 2 \alpha_2^d + \log 8)}{
                     \cosh ( 2 \pi \bar{\gamma}_k / \omega )}\\
C_1^{>}(k) &=& \cos^2 \left( \beta_1^d \right) \ \hspace{0.6cm} \ 
C_2^{>}(k) = \cos \left( \beta_2^d \right).
\end{eqnarray}

\begin{figure}[t]
\centerline{\epsfxsize=7.5cm{\epsfbox{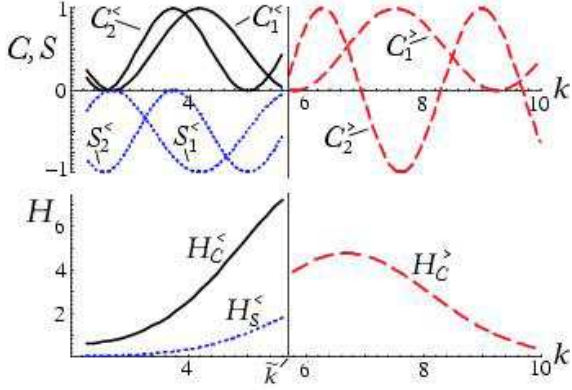}}} 
\caption{\label{fig:10} 
(Color online)
Main sinusoidal (top) and hyperbolic (bottom) components of the 
$Q_d(k,\Gamma^*)$ functions displayed in Fig.~\ref{fig:9}.
Their combination through equations (\ref{eq:QD<}) and (\ref{eq:QD>})
(for $k<\tilde{k}$ and $k>\tilde{k}$, respectively) determines the
amplitude envelope structure observed in Fig.~\ref{fig:9}.
}
\end{figure}

Figure \ref{fig:10} plots the $H$, $C$ and $S$ functions given above.
After close examination, one finds that the structure of the envelope
of $Q_d(k)$ is more complicated than that of $Q_a(k)$ mainly
because of the interplay between the oscillating $C$ and $S$ terms.
Indeed, the hyperbolic $H$ terms behave similarly to the $Q_a(k)$
case, presenting only one local maximum, and are therefore not
directly related to the appearance of multiple extrema in the
envelope.
For example, at $k \approx 5$ both $H_C^{<}(k)$ and $H_S^{<}(k)$ grow
with $k$ but the envelope of $Q_d(k)$ decreases, mainly because of the
oscillations of the $C_1^{<} C_2^{<}$ product.
Note that the change in the number of integration regions at
$\tilde{k}$ is not essential either for obtaining multiple extrema:
the combination of the oscillations of the $C$ and $S$ functions are
able to produce additional extrema even beyond their corresponding
domains.
Furthermore, in various tested cases with different fluid parameters
and forcing functions we have found no clear correlation between the
changes in the number of integration regions and the shape of the
neutral stability curves.
%

%
We now find analytic expressions that describe the envelope of the
resonance tongues for any forcing function with only two extrema per
cycle.
The neutral stability criterion in these cases is equivalent to
setting $Q_a(k,\Gamma)=1$ in expression (\ref{eq:QA}).
By dropping the oscillatory factor $\cos(\beta^a)$ in (\ref{eq:QA})
and using the high dissipation of the lubrication regime to neglect
the $\log 2$ term (when compared to $\alpha^a$ which, for the
parameters used in this section, is evaluated as $\alpha^a \approx 2
\pi \bar{\gamma}_k / \omega \approx 300$), we find that
\begin{equation}
\alpha^a \approx \frac{2 \pi \bar{\gamma}_k}{\omega}
\end{equation}
at the envelope. 
Using the definitions of $\alpha^a$ and $\bar{\gamma}_k$, this
condition can be rewritten as
\begin{equation}
\label{eq:IntChi}
\int_{V(x)>E} \sqrt{ | 1 - \chi + 
\Gamma_k\left( \chi \right) f(x) | } d x = 
2 \pi \sqrt{\chi},
\end{equation}
where the integration is carried out over the $V(x)>E$ region and the
algebraic function $\chi( k h )$ is given by
\begin{equation}
\label{def:Chi}
\chi(y) = \frac{\kappa_1 y^3}{1 + \kappa_2 y^2} B_1^2(y) B_2(y),
\end{equation}
with
\begin{eqnarray}
\kappa_1 = \frac{\nu^2}{g h^3} \ \, \, \mbox{   and   } \, \,
\kappa_2 = \frac{\sigma}{g \rho h^2}.
\end{eqnarray}
Equation (\ref{eq:IntChi}) provides an implicit expression for
$\Gamma_k(\chi)$ at the envelope.
Using this result and the definition in (\ref{def:Gammak}), we find
that the shape of the envelope of the resonance tongues under the
current approximations is described by the function
\begin{equation}
\label{eq:GammaE}
\Gamma_e(k h) = 
\left[ 1 + \kappa_2 k^2 h^2 \right] \Gamma_k \left(\chi(k h) \right).
\end{equation}
Unfortunately, there appears to be no simple way to extract the
properties of $\Gamma_e(k h)$ without further specifying $\kappa_1$,
$\kappa_2$ and $f(x)$.
However, we have observed for all tested cases that if $f(x)$ has only
two extrema per cycle, $\Gamma_e(k h)$ has only one minimum.
While the validity of this statement for all cases is a conjecture
that would require a proof which is beyond the scope of this paper, we
consider below two simple examples where analytic progress can be
made.

For square forcing (where $f(\omega t) = 1$ during half of the
period and $f(\omega t) = -1$ during the other half), the conjecture
can be proved as follows.
First, we find the solution of (\ref{eq:IntChi})
\begin{equation}
\label{eq:GammaChi}
\Gamma_k(\chi) = 3 \chi + 1.
\end{equation}
Then, we substitute this result into equation (\ref{eq:GammaE}) to
obtain an explicit expression for the envelope of the resonance
tongues
\begin{equation}
\Gamma_e^{\mathrm{sq}}(k h) = 
   3 \kappa_1 k^3 h^3 B_1^2 B_2 + \kappa_2 k^2 h^2 + 1.
\end{equation}
While the specific form of $\Gamma_e^{\mathrm{sq}}$ depends on the
parameters $\kappa_1$ and $\kappa_2$, its extrema can be readily
computed by using $\partial_k \Gamma_e^{\mathrm{sq}} = 0$. We find
that they are located at the intersection of the functions
$r(y) = - 3 \, \partial_y [ y^3  B_1^2(y) B_2(y) ]$
and
$s(y) = 2 \kappa_2 y / \kappa_1$.
Given that $r(y)$ does not depend on any parameters, it can be
evaluated numerically without loss of generality.
We find that it decreases monotonically, intersecting
the $r = 0$ axis at $y^* \approx 1.479$.
Using this result and the fact that $s(y)$ is a linearly increasing
function, it is easy to see that $\Gamma_e^{\mathrm{sq}}(k h)$ can
have only one minimum (which must be located at $k \leq y^*/h$).

For triangular forcing, (where $f(\omega t)$ is a linear function
that increases during half of the period and a decreases during the
other half), the analytical calculation becomes much harder.
The solution for $\Gamma_k$ is given by the real root of the cubic
equation
\begin{equation}
\label{eq:triang}
\left( \Gamma_k + \chi - 1 \right)^3 = 9 \Gamma_k^2 \chi.
\end{equation}
It has a more complicated structure than (\ref{eq:GammaChi}), which
renders the use of the techniques developed for the square forcing
case impossible.
In the current analysis we will therefore content ourselves with
scanning the parameter space numerically to show that, for a wide
range of systems with triangular forcing, the envelope of the
resonance tongues has only one minimum.
In order to do this, we first note that the problem now depends on
only two nondimensional parameters: $\kappa_1$ and $\kappa_2$.
We also note that we can write the analytic solution of
(\ref{eq:triang}) and use (\ref{def:Chi}) and (\ref{eq:GammaE}) to
obtain a (very long) explicit algebraic expression for the envelope of
the resonance tongues, which we label $\Gamma^{\mathrm{tri}}_e(k h)$
but do not reproduce here because of its length.
By evaluating $\partial_k^2 \Gamma^{\mathrm{tri}}_e(k h)$ at $10^3$
points between $k=0$ and $k$-values that reach an asymptotic regime,
using approximately $10^4$ different (logarithmically spaced)
combinations of the parameters $\kappa_1 \in [10^{-6}, 10^{1}]$ and
$\kappa_2 \in [10^{-5}, 10^{3}]$, we find that
$\Gamma^{\mathrm{tri}}_e(k h)$ is always a smooth function with
positive concavity.
This strongly suggests that $\Gamma^{\mathrm{tri}}_e(k h)$ has only
one minimum and that the conjecture also holds for triangular
forcings.

Finally, for a sinusoidal forcing $f(\omega t) \propto \cos(\omega t)$
one can only express $\Gamma_k(\chi)$ in terms of an integral equation
which cannot be explicitly solved.
The work in \cite{Cerda1,Cerda2}, however, shows that 
$\Gamma^{\mathrm{sin}}_e(k h)$ again appears to have only one minimum
for any combination of parameters.

The results presented above relate the shape of the forcing function
to that of the envelope of the resonance tongues.
In particular, they support the conjecture that only a forcing with
more than two extrema per cycle can generate a tongue envelope that
has more than one minimum.
A full proof of this conjecture would be of interest not only as a
mathematical result, but also as a guide for engineering surface
patterns.
It would imply, for example, that only forcing functions that have
this characteristic can display bicritical points involving
non-contiguous resonance tongues.
%


\section{Discussion and conclusions}
\label{sec:DiscAndConcl}

We have presented a new approach for studying the effect of the shape
of the forcing function on the Faraday linear surface wave
instabilities.
Through a numerical, experimental and analytic investigation, we have
established a relation between the number of extrema in the forcing
function and the number of minima that can appear in the envelope of
the resonance tongues.
This approach does not rely on a multi-frequency description of the
forcing function. It therefore allows us to consider forcings that
cannot be defined by the superposition of a few sinusoidal terms, but
that can excite surface wave instabilities in new ways that could lead
to a greater control of the surface patterns.

The analysis that we have carried out provides new insights for
understanding the effects of the energy feeding mechanism in pattern
forming systems.
Indeed, we use the lubrication approximation to reduce the system to
one degree of freedom and then apply the WKB method, which neglects
the fast oscillations by integrating their net effect over the
different forcing segments.
By doing this, we achieve a description that is somehow similar to the
simple mechanical analogies (with balls, springs and pendula) that are
used in reduced dimensionality models of parametric resonance.
In this context, it would be interesting to try to relate the
simplified dynamics that the WKB calculations furnish for each
wavenumber to the forcing strength required to reach its corresponding
instability threshold.
Furthermore, it may be possible to follow a similar approach to study
the effects of the forcing mechanism in other fluid regimes or even in
a different system, such as the granular Faraday experiments where
strongly non-sinusoidal forcings is the norm \cite{Paul1}.

From an analytical perspective, various additional connections between
the forcing shape and the resonance tongues could be obtained by
developing the implicit relations established here.
We expect to be able to achieve this by adequately choosing a reduced
set of forcing functions and using the right approximations.
Obtaining these additional connections could lead to a better
understanding of the inverse problem, in which the forcing function
would be tailored to achieve a given instability.

From an experimental perspective, the lubrication regime in which our
analytic results are obtained has not yet been widely explored.
This is not due to any fundamental limitation but rather to technical
difficulties, mainly in achieving high enough accelerations at low
frequencies and having a large enough container for the surface
patterns to develop.
However, given that we obtain good analytical approximations in this
regime, we hope that new experiments will explore this regime.
This, together with an extension of our analysis to consider nonlinear
effects, would allow an exploration of the patterns that can be formed
by the linear instabilities achieved through the forcing function
control.

\section{Acknowledgments}

MS acknowledges partial support from NASA Grant NAG3-2364 and NSF 
Grant DMS-0309667.


\bibliography{journal}

\end{document}